\documentclass[12pt]{article}
\usepackage{amsmath}
\usepackage{graphicx}
\usepackage{amsfonts}
\usepackage{amssymb}
\begin{document}

\title{Grand unification using a generalized Yang-Mills theory}
\author{M. Chaves\\\textit{Escuela de F\'{\i}sica, Universidad de Costa Rica}\\\textit{San Jos\'e, Costa Rica}\\\textit{E-mail: mchaves@cariari.ucr.ac.cr}\\and \\H. Morales\\\textit{Department of Physics \& Astronomy}\\\textit{University of Kentucky}\\\textit{Lexington, Kentucky 40506-0055}\\\textit{E-mail: hmorales@pa.uky.edu}}
\date{November 8, 1999}
\maketitle
\begin{abstract}
Generalized Yang-Mills theories have a covariant derivative that employs both
scalar and vector bosons. Here we show how grand unified theories of the
electroweak and strong interactions can be constructed with them. In
particular the SU(5) GUT can be obtained from SU(6) with SU(5)xU(1) as a
maximal subgroup. The choice of maximal subgroup also determines the chiral
structure of the theory. The resulting Lagrangian has only two terms, and only
two irreducible representations are needed, one for fermions and another for bosons.
\end{abstract}

\section{Introduction}

\noindent The idea behind Generalized Yang-Mills theories$^{\text{1-3}}$
(GYMTs) is as follows: In the usual Yang-Mills theory gauge invariance is
assured through the demand that vector gauge fields $A_{\mu}$ transform as
$A_{\mu}\rightarrow UA_{\mu}U^{-1}-(\partial_{\mu}U)U^{-1}.$ In a GYMT scalar
boson fields are considered to be also acceptable as gauge fields, with a
transformation law similar to the one just shown. To be able to sum together
scalar and vector bosons, it is necessary to express all terms of the theory,
both bosonic and fermionic, in the spinorial representation. Technical details
on this procedure are left for next section. As will be seen, a gauge
invariant theory is obtained, which we call a GYMT. In the usual Yang-Mills
theories gauge bosons are placed in the adjoint representation; in GYMTs
whatever Higgs fields are needed are placed there, too. The total number of
bosons, vector plus scalar, has to equal the number of group generators, a
rather stringent requirement on their numbers. The fermions are placed in
another irreducible representation (irrep) of mixed chirality. The structure
of the adjoint gauge field matrix is%
\begin{equation}
\left(
\begin{array}
[c]{cc}%
\text{vector bosons} & \text{scalar bosons}\\
\text{scalar bosons} & \text{vector bosons}%
\end{array}
\right)  ,\label{structureb}%
\end{equation}
and that of the fermion field matrix is:%
\begin{equation}
\left(
\begin{array}
[c]{cc}%
\text{left fermions} & \text{right fermions}\\
\text{right fermions} & \text{left fermions}%
\end{array}
\right)  .\label{structuref}%
\end{equation}
The possible dimensions of the blocks that make up the matrices are determined
by the maximal subgroups of the gauge group. If we assume it to be $SU(6),$
then there are three possible maximal subgroups, $SU(3)\times SU(3)\times
U(1),$ $SU(4)\times SU(2)\times U(1),$ and $SU(5)\times U(1).$ In this case,
for example, the upper left blocks for both matrices would be a square of side
3, 4, and 5, respectively.

The original inspiration for our work is an old idea due to Fairlie and
Ne'eman,$^{4}$ of using $SU(2/1)$ as a unification group for the
Glashow-Weinberg-Salam (GWS) model, and putting the Higgs fields in the
adjoint along with the vector fields. However, in spite of much effort by
these and other authors, no success was achieved at the time in these matters.
The belief that the gauge group had to be graded$^{1,2}$ lead to insurmountable difficulties.

It is possible to construct a grand unified theory (GUT) of the strong and
electroweak forces using a GYMT.$^{3}$ The immediate benefit of this procedure
is simplicity. There are only two terms in the Lagrangian, whereas in the
usual GUTs there are several. As mentioned, only two irreps are needed, as
compared with one or two for the fermions, at least two for the Higgs fields
and one for the gauge vector fields, in the case of a usual GUT.

In high energy physics the word ``unification'' has a very precise meaning:
one selects a large Lie group, which contains in a maximal subgroup other
small Lie groups that represent the forces that are to be ``unified''. Then,
through the process of spontaneous symmetry breaking brought about by nonzero
vacuum expectation values of Higgs fields, the smaller groups can be recovered
and with them the low energy phenomenology. In the case of GYMT grand
unification the same process is occurring insofar as one must choose a large
Lie group so that it can contain the smaller phenomenological groups, but,
besides this process, there is another kind of ``unification'', brought about
by putting together in one irrep vector and scalar bosons (which in turns
permits to put together in the same irrep left and right fermions). To go from
this second kind of ``unification'' to the smaller groups, it is necessary to
restrict the gauge group to its maximal subgroup in order to obtain a
Yang-Mills theory and be in a more familiar terrain, so to speak.

The ``generalized gauge transformation'' leaves invariant the Lagrangian of
the theory. However, this does not mean that \emph{separate} parts of the
Lagrangian are necessarily kept invariant if they are transformed
individually. As an example, consider the fermion matrix (\ref{structuref}),
whose chiral structure is not left invariant by the most general ``generalized
gauge transformation''. But the maximal subgroup \emph{does} leave its chiral
structure invariant. In fact, the different maximal subgroups are the possible
Yang-Mills theories derivable from a particular GYMT.

A very general ``generalized gauge transformation'' is not going to help us
recover low energy phenomenology, since it leads to expressions we are not
accustomed to. The procedure is to obtain the Yang-Mills theory from the GYMT
first by restraining the gauge group to a maximal subgroup; then, through the
usual spontaneous symmetry breaking processes, obtain the small
phenomenological groups. We shall come back to this topic in last section, in
relation to the example covered in this paper, $SU(6)$, when the discussion is
not so abstract.

\section{GYMTs}

The first thing we have to do is learn how to write covariant derivatives in
the spinorial representation. Let $S$ be an element of the spinor
representation of the Lorentz group, so that, if $\psi$ is a spinor, then it
transforms as $\psi\rightarrow S\psi.$ Then, due to the homomorphism that
exists between the vector and spinor representations of the Lorentz group, we
have that $A\!\!\!/\rightarrow SA\!\!\!/S^{-1}$ for a vector $A_{\mu}$ that
has been contracted with the Dirac matrices. It was this homomorphism that
allowed Dirac to write a spinorial equation that included the vector
electromagnetic field. We have the following

\textbf{Theorem. }\emph{Let }$D_{\mu}=\partial_{\mu}+B_{\mu}$\emph{, where
}$B_{\mu}$\emph{\ is some vector field. Then: }
\begin{equation}
(\partial_{\mu}B_{\nu}-\partial_{\nu}B_{\mu})(\partial^{\mu}B^{\nu}%
-\partial^{\nu}B^{\mu})=\frac18\mathrm{Tr}^{2}D\!\!\!\!/^{\,2}-\frac
12\mathrm{Tr}D\!\!\!\!/^{\,4}, \label{teo_1}%
\end{equation}
\emph{where the traces are to be taken over the Dirac matrices.}

We are not going to prove this theorem. Instead, we remit the reader to
Ref.~1, where the proof is done in detail. Consider now a Yang-Mills theory
that is invariant under a Lie group with $N$ generators. The fermion or matter
sector of the non-Abelian Lagrangian has the form $\bar{\psi}iD\!\!\!\!/\psi$,
where $D_{\mu}$ is a covariant derivative chosen to maintain gauge invariance,
and $\psi$ is a chiral fermion multiplet. This multiplet transforms as
$\psi\rightarrow U\psi$, where $U=U(x)$ is an element of the fundamental
representation of the group. The covariant derivative is $D_{\mu}%
=\partial_{\mu}+A_{\mu}$, where $A_{\mu}=igA_{\mu}^{a}(x)T^{a}$ is an element
of the Lie algebra and $g$ is a coupling constant. We are assuming here that
the matrices $\{T^{a}\},$ $a=1,\ldots,N,$ constitute a representation of the
group generators. Gauge invariance of the matter term is assured if
\begin{equation}
A_{\mu}\rightarrow UA_{\mu}U^{-1}-(\partial_{\mu}U)U^{-1}\,{,} \label{vec_2}%
\end{equation}
or, what is the same,
\begin{equation}
A\!\!\!/\rightarrow UA\!\!\!/U^{-1}-(\partial\!\!\!/U)U^{-1}. \label{vec_3}%
\end{equation}

To construct the GYMT we generalize the transformation of the gauge field as
follows: to every generator in the Lie group we choose one gauge field that
can be either vector or scalar. Suppose there are $N$ generators in the Lie
group; we choose $N_{V}$ generators to be associated with an equal number of
vector gauge fields and the other $N_{S}$ to be associated with an equal
number of scalar gauge fields. Naturally $N_{V}+N_{S}=N$. The choice has to be
made so that the adjoint can be decomposed in blocks made up solely of vector
or scalar fields (with the exception of the fields along the diagonal, that
can be either), as shown in (\ref{structureb}), with the block size given by
the dimension of the subgroups forming the maximal subgroup. We define the
generalized covariant derivative $D$ by taking each one of the generators and
multiplying it by one of its associated gauge fields and summing them
together. The result is%
\begin{equation}
D\equiv\partial\!\!\!/+A\!\!\!/+\Phi\label{der_na}%
\end{equation}
where $A\!\!\!/=\gamma^{\mu}A_{\mu}=ig\gamma^{\mu}A_{\mu}^{a}T^{a},$
$a=1,\ldots,N_{V},$ and $\Phi=\gamma^{5}\varphi=-g\gamma^{5}\varphi^{b}T^{b},$
$b=N_{V}+1,\ldots,N.$

We now define the transformation for the gauge fields to be
\begin{equation}
A\!\!\!/+\Phi\rightarrow U(A\!\!\!/+\Phi)U^{-1}-(\partial\!\!\!/U)U^{-1}%
,\label{vec_4}%
\end{equation}
from which one can conclude that $D\rightarrow UDU^{-1}.$ The following
Lagrangian is built based on the conditions that it contain only matter fields
and covariant derivatives, and possess both Lorentz and gauge invariance:%
\begin{equation}
\mathcal{L}_{NA}=\bar{\psi}iD\psi+\frac{1}{2g^{2}}\widetilde{\mathrm{Tr}%
}\left(  \frac{1}{8}\mathrm{Tr}^{2}D^{\,2}-\frac{1}{2}\mathrm{Tr}%
D^{\,4}\right)  ,\label{lag_na}%
\end{equation}
where the trace with the tilde is over the gauge (or Lie group) matrices and
the one without it is over matrices of the spinorial representation of the
Lorentz group. (For some fermion irreps a small modification is necessary. See
Section 5.)

Although we have constructed this non-Abelian Lagrangian based only on the
conditions just mentioned, its expansion into component fields results in
expressions that are traditional in Yang-Mills theories. The expansion is done
in detail in Ref.~1, and results in%
\begin{align}
\mathcal{L}_{NA}  &  =\bar{\psi}i(\partial\!\!\!/+A\!\!\!/)\psi-g\bar{\psi
}i\gamma^{5}\varphi^{b}T^{b}\psi+\frac{1}{2g^{2}}\widetilde{\mathrm{Tr}%
}\left(  \partial_{\mu}A_{\nu}-\partial_{\nu}A_{\mu}+[A_{\mu},A_{\nu}]\right)
^{2}\nonumber\\
&  +\frac{1}{g^{2}}\widetilde{\mathrm{Tr}}\left(  \partial_{\mu}%
\varphi+[A_{\mu},\varphi]\right)  ^{2}. \label{expanded}%
\end{align}
The reader will recognize familiar structures: the first term on the right
looks like the usual matter term of a gauge theory, the second like a Yukawa
term, the third like the kinetic energy of vector bosons in a Yang-Mills
theory and the fourth like the gauge-invariant kinetic energy of scalar bosons
in the non-Abelian adjoint representation.

\section{A case study: the GWS model}

The GWS model can be written$^{1}$ as a GYMT using the group $U(3).$ It is
necessary to use this group and not $SU(3),$ as we shall see. Referring to
(\ref{expanded}), notice that one can infer, from the terms $\bar{\psi
}iA\!\!\!/\psi$ and $\frac{1}{2g^{2}}\widetilde{\mathrm{Tr}}\left(  [A_{\mu
},A_{\nu}]\right)  ^{2},$ that the quantum numbers of the particles of a GYMT
are the same and can be calculated in the same ways as if it were a
conventional Yang-Mills theory. In particular, notice that a maximal subgroup,
that would only keep diagonal terms in (\ref{structureb}), would have the same
diagonal generators (and thus the same quantum numbers), as the full group.

Originally, in Ref.~1, the authors calculated the quantum numbers directly
from the diagonal generators of the group, but in this paper we use, when
possible, group-theoretical techniques. We follow the conventions used in
Slansky's handbook ``Group theory for unified model building''$^{5}$ except
that we shall write hypercharges as subindices and not between parenthesis. In
particular we use the relation $Q=T_{3}+\frac{1}{2}Y,$ with a positive sign
for the hypercharge term as in this reference and not as we have done in
previous papers, with a minus sign. The group theory results we use are for
the most part listed in Slansky's Table 58 on branching rules for maximal subgroups.

If we try to construct a GYMT of the GWS model using $SU(3),$ the bosons are
placed in the adjoint $8.$ This irrep has, under the maximal subgroup
$SU(3)\supset SU_{I}(2)\times U_{Y}(1),$ the branching rule $8=1_{0}%
+2_{3}+\bar{2}_{-3}+3_{0};$ that is, the $8$ is the direct sum of an isospin
singlet with $Y=0,$ a doublet with $Y=3,$ and so on. The gauge field matrix
$\mathcal{G}_{3},$ that appears in the covariant derivative $D_{3}%
=\partial\!\!\!/+\mathcal{G}_{3}$ would, omitting all numerical factors, look
like%
\begin{equation}
\mathcal{G}_{3}=\!\left(
\begin{array}
[c]{ccc}%
B\!\!\!\!/+A\!\!\!/_{3} & A\!\!\!/_{+} & \varphi_{1}\\
A\!\!\!/_{-} & B\!\!\!\!/-A\!\!\!/_{3} & \varphi_{2}\\
\varphi_{1}^{\ast} & \varphi_{2}^{\ast} & -2B\!\!\!\!/
\end{array}
\right)  .\label{gauge3}%
\end{equation}
Algebraically, to obtain the quantum number of a field one must take the
commutator of the field with the diagonal generator associated with that
quantum number. Thus the hypercharges of the gauge fields in $\mathcal{G}_{3}$
are given by the coefficients of the fields in $[\lambda_{8},\mathcal{G}%
_{3}],$ where $\lambda_{8}\propto\operatorname{diag}(1,1,-2)$. For the case of
$\varphi_{1}$ or $\varphi_{2}$ one gets $Y=3.$ Alternatively, one can obtain
this result directly and immediately from the branching rule for the $8,$
since the Higgs fields belong in the $2_{3}$. The problem is that the GWS
Higgs field has $Y=1,$ not $Y=3$.

It is not possible to construct a GYMT for the GWS using $SU(3),$ but with
$U(3)$ it is possible to do it. For in this group there is the additional
generator $\lambda_{9}\propto\operatorname{diag}(1,1,1),$ and it is possible
use it, along with $\lambda_{8},$ to form an alternative representation with
generators $\lambda_{10}\propto\operatorname{diag}(1,1,-1)$ and $\lambda
_{0}\propto\operatorname{diag}(1,1,2),$ which can be expressed as two
orthogonal linear combinations of $\lambda_{8}$ and $\lambda_{9}$. We define
$\lambda_{0}$ to be the correct hypercharge generator, and with it the value
$Y=1$ is obtained for the hypercharge of the Higgs. With this hypercharge we
have a natural choice for the fermions of the GWS model:%
\begin{equation}
\psi=\left(
\begin{array}
[c]{c}%
e_{R}^{c}\\
\nu_{R}^{c}\\
e_{L}^{c}%
\end{array}
\right)  .\label{chiral}%
\end{equation}
This choice of hypercharge gives all the particles of the GWS model with their
correct quantum numbers. We do not obtain the Higgs potential energy density
$V(\varphi)$ from GYMTs.

It seems peculiar that one has to go to an unusual representation of $U(3)$ to
be able to construct the GYMT. The explanation for this peculiarity becomes
evident if one considers grand unified GYMTs. The group $SU(6)$ has the
maximal subgroup $SU(6)\supset SU(3)\times SU(3)\times U(1),$ which, in a way,
seems the logical one to use since it contains two $SU(3)$, one for the GWS
model and another for chromodynamics$.$ We actually attempted grand
unification first this way$^{3}$ but we were forced to use an unusual
representation of the generators of $SU(6)$ to obtain the correct quantum
numbers of the particles of the Standard Model. We also overlooked, in part
due to the difficult algebra, that it is necessary to use a mixed field for
the hypercharge generator, as we shall explain next section. Here we shall use
a different maximal subgroup, $SU(6)\supset SU(5)\times U(1).$ With it is not
necessary to go to a new representations to get the correct quantum numbers
for the vector bosons, Higgs bosons, or fermions of the Standard Model, and
the relation between the $SU(5)$ GUT and $SU(6)$ GYMT becomes wonderfully transparent.

\section{The bosons of the $SU(6)\supset SU(5)\times U(1)$ GYMT GUT}

\noindent The GYMT based on $SU(6)$ will use the adjoint $35$ for the bosons
and the $15$ for the fermions. This last irrep is the antisymmetric product of
two $6$'s$,$ the $6$ being the fundamental representation. To obtain the
particle spectrum of the theory we use the maximal subgroup $SU(5)\times
U(1).$ The branching rule for the 35 is $35=1_{0}+5_{6}+\bar{5}_{-6}+24_{0}.$
The $5$ is the fundamental of $SU(5),$ and the $24$ is the adjoint.
Schematically, the gauge fields look like this:%
\begin{equation}
\mathcal{G}_{6}=\left(
\begin{array}
[c]{cc}%
24 & 5\\
\bar{5} & 1
\end{array}
\right)  =\left(
\begin{array}
[c]{cc}%
\mathcal{G}_{5}+\Omega & \varphi\\
\varphi^{\dagger} & -5\Omega
\end{array}
\right)  .\label{gauge6}%
\end{equation}
Here $\mathcal{G}_{5}$ transforms as the adjoint $24$ of $SU(5);$ $\varphi,$
the GWS Higgs field, as the $5$; $\varphi^{\dagger}$ as the $\bar{5};$ and
$\Omega$ as the singlet, all according to the branching rule $35=1_{0}%
+5_{6}+\bar{5}_{-6}+24_{0}$. These irreps of the branching of the $35$ under
$SU(6)\supset SU(5)\times U(1)$ take the place of the irreps of the normal
$SU(5)$ GUT, with the difference while the GUT Higgs in $SU(5)$ transforms as
another $24$ here it does it as the singlet $1_{0}$. While in the usual GUT
theory each irrep is arbitrarily chosen, in GYMTs they are basically given by
the choice of the gauge group and its maximal subgroup.

The hypercharge generator for the $SU(5)$ GUT is of the form
$\operatorname{diag}(-2,-2,-2,3,3),$ so for our $SU(6)$ case%
\begin{equation}
Y\propto\operatorname{diag}(-2,-2,-2,3,3,0),\label{Y}%
\end{equation}
and we assign to it the vector boson $B\!\!\!\!/$, in a way similar to the way
it is done in $SU(5)$ GUT. (We shall see that actually one has to assign to it
a mixture of $B\!\!\!\!/$ and $\Omega$, this last being a scalar we proceed to
introduce.) From the structure of the maximal subgroup $SU(5)\times U(1)$ one
knows that there is a generator
\begin{equation}
Z\propto\operatorname{diag}(1,1,1,1,1,-5).
\end{equation}
We shall give the name \emph{ultracharge} to the quantum number associated
with the $Z$ diagonal generator. It is necessary to assign a scalar boson, we
call $\Omega,$ to this generator, since we cannot assign a vector boson to it
because in that case it should have had already been observed, being massless.
There is no problem assigning a scalar field, for, as can be seen in
(\ref{expanded}), the terms that would couple scalar bosons among themselves
are missing. These terms drop out from Lagrangian (\ref{lag_na}) for algebraic
reasons. This scalar $\Omega$ has to be the same one that generates the
large-scale GUTs masses to the leptoquark bosons, since there are no more
bosons available. It should have a very large mass, since Higgs bosons are
usually assumed to have masses of the order of their vacuum expectation values.

Just as in a normal GUT, phenomenology is obtained through the maximal
subgroup $SU(5)\supset SU_{I}(2)\times SU_{C}(3)\times U_{Y}(1).$ The
branching of the adjoint is%
\begin{equation}
24=(1,1)_{0}+(3,1)_{0}+(\bar{2},3)_{-5}+(2,\bar{3})_{5}+(1,8)_{0},\label{24}%
\end{equation}
where each irrep is given in the form $(m,n)_{Y},$ where $m$ is in $SU_{I}(2)$
and $n$ in $SU_{C}(3).$ The bosons contained in $\mathcal{G}_{5}$ are the
usual four vector bosons of the GWS model, the eight gluons, and the twelve
leptoquark bosons. If we add to these the five complex Higgs fields $\varphi$
and the scalar singlet $\Omega$ we obtain the 35 fields that make up the GYMT adjoint.

While this arrangement seems nicely precise, there is one serious difficulty
with it. As can be seen from the branching rule for the 35, the ultracharge of all the bosons is
zero, so that, in particular, the leptoquarks do not acquire a very large
mass. We propose a simple solution to this problem that is reminiscent of the
special representation of the generators of the $U(3)$ group that was
necessary to introduce to be able to express the GWS model as a GYMT. It
consists in assigning to the hypercharge generator a field $\Xi$ that is the
mixture of two fields that have already been introduced:%
\begin{equation}
\Xi=B\!\!\!\!/+\gamma^{5}\Omega.\label{mixture}%
\end{equation}
There is no change in the field assigned to the ultracharge generator; it is
still the $\Omega.$ With this assignment the leptoquarks (and only them)
couple to the $\Omega,$ as can be seen in (\ref{24}) from the hypercharge
assignments, so they obtain a very large mass.

In next section we study the fermion sector of the model. The reader may
perhaps find it unlikely that the scheme we have introduced could work with
fermions, first, because the mixed field $\Xi$ would apparently imply the
coupling of $\Omega$ to fermions, giving them a mass on the scale of grand
unification, and, second, that, while no fermion must couple to the $\Omega,$
some of them must necessarily couple to the $\varphi$ bosons to reproduce the
GWS model. Actually the fermion terms appear with a serendipitous combination
of chiralities that gives all the correct results.

\section{The fermions of the $SU(6)\supset SU(5)\times U(1)$ GYMT GUT}

The fermions in the $SU(6)$ GYMT GUT are placed in the antisymmetric $15.$ The
branching rule for this irrep for the maximal subgroup $SU(5)\times U(1)$ is
$15=5_{-4}+10_{2},$ where the subindex is the ultracharge. In turn the
branchings of these two irreps are $5=(1,3)_{-2}+(2,1)_{3}$ and $10=(2,3)_{1}%
+(1,\bar{3})_{-4}+(1,1)_{6}$ where, as before, the irreps are expressed in the
form $(m,n)_{Y},$ where $m$ is in $SU_{I}(2)$ and $n$ in $SU_{C}(3).$ These
two irreps also appear in normal $SU(5),$ and on the basis of the isospin,
hypercharge, and color quantum numbers their particle content is
$5\rightarrow(d^{1},d^{2},d^{3},e^{c},\nu^{c})_{R}$ and $10\rightarrow
(u^{1},d^{1},u^{2},d^{2},u^{3},d^{3},u^{1c},u^{2c},u^{3c},e^{c})_{L}.$ The
quark colors have been denoted 1, 2 and 3. It is possible to place these two
irreps in a single one, using larger unification groups, but then the two
irreps should be $\bar{5}$ and $10,$ charge conjugating the $5,$ in order to
have the same chiralities. However, in a GYMT GUT just the opposite happens:
the two irreps must necessarily have different chiralities to be included in a
larger irrep, so that, denoting the chirality of the irrep by a subindex, the
$5_{R}$ and the $10_{L}$ are just what we need.

The fermion irreps fit into the antisymmetric $15$ in the following way:%
\begin{equation}
\psi=\left(
\begin{array}
[c]{cc}%
10_{L} & 5_{R}\\
-5_{R}^{T} & 0
\end{array}
\right)  . \label{15}%
\end{equation}
This field transforms as the antisymmetrized Kronecker product of two
fundamentals of $SU(6)$, so that, if $U$ is an element of the fundamental
representation$,$ then under a group transformation $\psi\rightarrow U\psi
U^{T}$ and $\bar{\psi}\rightarrow U^{\ast}\bar{\psi}U^{-1}.$ Furthermore, the
adjoint is the product of the fundamental times its conjugate, $\mathcal{G}%
_{6}\rightarrow U\mathcal{G}_{6}U^{-1},$ so that the correct invariant fermion
term in this GYMT Lagrangian would need a trace over the gauge group:%
\begin{equation}
\widetilde{\mathrm{Tr}}\left(  \bar{\psi}i(\partial\!\!\!/+\mathcal{G}%
_{6})\psi\right)  . \label{fermiontrace}%
\end{equation}

It is not necessary to prove that the $5$ and $10$ predict the correct fermion
spectrum, since they are the same irreps that are used in the usual $SU(5)$
GUT. What we do have to verify is that the mixed field $\Xi$ causes no
inconsistencies with the fermion phenomenology of the Standard Model.
Expanding (\ref{fermiontrace}) into all its components, one obtains three
types of terms:

\begin{enumerate}
\item  Type $\bar{\chi}_{R}C\!\!\!\!/\,\theta_{R}$ or $\bar{\chi}%
_{L}C\!\!\!\!/\,\theta_{L},$ where $C_{\mu}$ is any one of the vector bosons
of the Standard Model. (The definition of $\bar{\chi}_{R}$ is $(\chi
_{R}^{\dagger}\gamma^{0})_{L},$ so it is a spinor with left chirality.) These
terms represent the usual couplings of the Standard Model between fermions and
vector bosons, including leptoquarks.

\item  Type $\bar{\chi}_{L}\varphi\,\theta_{R}$ or $\bar{\chi}_{R}%
\varphi\,\theta_{L},$where $\varphi$ is the Higgs that gives mass to the
$W^{\pm}$ and $Z^{0},$ the carriers of the weak interaction.

\item  Type $\bar{\chi}_{R}\Omega\theta_{R}$ or $\bar{\chi}_{L}\Omega
\theta_{L},$ that are zero due to the opposite chiralities. As we said, it was
important that these terms cancelled.
\end{enumerate}

\noindent Fortunately, in each case the correct combination of chiral spinors
has appeared. Any variation would have been phenomenologically unacceptable.

The couplings with the vector bosons are the same as those predicted by the
$SU(5)$ GUT. There is a term present in the $SU(5)$ GUT that is missing in
this $SU(6)\supset SU(5)\times U(1)$ GYMT. The $SU(5)$ gauge invariance allows
two Yukawa terms that couple fermions $5_{R}$ and the $10_{L}$ with the Higgs
boson $5.$ The two invariants are $\overline{10}_{L}\times5\times5_{R}$ and
$\overline{10}_{L}^{c}\times5\times10_{L}=10_{L}\times5\times10_{L}.$ In this
GYMT there is only one term that involves fermions and it is given in
(\ref{fermiontrace}). The only $SU(6)$ gauge invariant is $\overline{15}%
\times35\times15,$ which contains only the first of the $SU(5)$ invariants.
Fortunately all the fermion kinetic energy terms are present in $\overline
{10}_{L}\times5\times5_{R},$ so what is missing is this GWS Higgs'
contribution to the mass of the right quark.

\section{Final comments}

\noindent We have defined GYMTs, and shown that is possible to construct a
satisfactory GUT with them. The group $SU(6)$ with the $SU(5)\times U(1)$
maximal subgroup gives results that are very simple to interpret. With
$SU(5)\times U(1)$ one obtains all the correct quantum numbers of the
particles of the Standard Model right away, without having to perform first a
rotation in root space, as it is the case using other maximal subgroups.
Furthermore, the relation between the $SU(5)$ GUT and the $SU(6)$ GYMT becomes
transparent. To put it bluntly, the first is the block expansion, as dictated
by the maximal subgroup, of the matrices of the second. Another important
point is that to the $SU(6)$ hypercharge generator a mixture of two fields
must be assigned to correctly obtain the Standard Model: the usual hypercharge
vector and the GUT Higgs. There are still 35 boson fields for the 35 $SU(6)$ irrep.

One advantage of this formulation of the Standard Model is that it only
requires two irreps, one for all the bosons and another for all the fermions.
Not only does the GWS Higgs, but also the GUT Higgs, fit in the $SU(6)$'s
$35.$ The Lagrangian of the GYMT has only two terms: the fermion kinetic
energy and a term in powers of $D.$

Notice that our motivation to restrict the symmetry to $SU(5)\times U(1)$ is
due to the need to obtain a theory one is familiar with, in this case the
underlying $SU(5)$ Yang-Mills structure. Then there are two additional
spontaneous symmetry breakings, one from $SU(5)$ to $SU(3)_{C}\times
SU_{I}(2)\times U_{Y}(1)$ and another from $SU_{I}(2)\times U_{Y}(1)$ to
$U_{EM}(1),$ but these are of the usual type, due to nonzero vacuum
expectation values of Higgs fields.

We still do not know how to include in a natural way in the model the Higgs
potential $V(\varphi,\Omega)$. What we have done is simply to assign a
GWS-scale vacuum expectation value to the colorless, chargeless, component of
the Higgs field $\varphi,$ and a GUT-scale to the Higgs field $\Omega,$ which
is a colorless and chargeless singlet. Incidentally, the right neutrino cannot
be included in a GYMT GUT using $SU(6).$ To include it would necessitate a
larger group. As was mentioned at the end of last section, the term that
couples the GWS Higgs with the up quark in the $SU(5)$ GUT seems to be missing
in this GYMT, so that this quarks does not get mass generation from $\varphi.$
However the fermionic term that is present does contain the up quark's kinetic
energy term.

Just as in Yang-Mills theories when the covariant derivative acts on a field
$X$ its gauge fields acquire certain coefficients called the ``charges'' of
each gauge field with respect to the field $X,$ in this generalization the
same thing has to be done. Thus, when $D$ acts on the leptonic triplet, its
gauge fields are going to be multiplied by constants $Q_{V}$ and $Q_{S},$ with
the result $D\psi=(\partial\!\!\!/+Q_{V}A\!\!\!/+Q_{S}\Phi)\psi.$ From our
knowledge of the Standard Model we conclude that $Q_{V}=1$ and that there are
three $Q_{S},$ one for each generation, with rather small values.

While it is the maximal subgroup that generates the phenomenology of a
GUT\ GYMT, there is an additional, powerful symmetry present: the gauge group
itself, $SU(6),$ in this case. The implications of this symmetry seem to us to
be a worthwhile subject of study for the future.\bigskip\ 

\noindent\textbf{References}

\noindent1. M. Chaves and H. Morales, Mod. Phys. Lett. \textbf{A13}, 2021 (1998).

\noindent2. M. Chaves, ``Some Mathematical Considerations about Generalized
Yang-Mills Theories'', in \emph{Photon and Poincare Group}, ed. V. V.
Dvoeglazov, (Nova Science Publishers, New York, 1998). This and the previous
reference contain fairly complete bibliographies.

\noindent3. M. Chaves and H. Morales, ``The Standard Model and the Generalized
Covariant Derivative'', in Proceedings of the International Workshop ``Lorentz
Group, CPT, and Neutrinos'', Universidad Aut\'{o}noma de Zacatecas,
M\'{e}xico, June 23-26, 1999, (World Scientific), to be published.

\noindent4. D. B. Fairlie, Phys. Lett. \textbf{B82}, 97 (1979); Y. Ne'eman,
Phys. Lett. \textbf{B81}, 190 (1979).

\noindent5. R. Slansky, Phys. Rep. \textbf{79}, 1 (1981).
\end{document}